\documentclass[twocolumn,preprintnumbers,amsmath,amssymb,prb]{revtex4}
\usepackage{graphicx}
\usepackage{fancyhdr}
\usepackage{dcolumn}
\usepackage{bm}
\usepackage{epstopdf}
\usepackage{caption}

 \usepackage[colorlinks,  linkcolor=red, anchorcolor=green, citecolor=blue]{hyperref}
  \usepackage{natbib}
  \usepackage[a4paper,left=1.55cm,top=2cm,right=1.55cm]{geometry}
\usepackage{amssymb}
\usepackage{amsmath}
\usepackage[normalem]{ulem}
\usepackage[dvips]{color}

\begin{document}
\title[\texttt{achemso} demonstration]
{Fast Water Channeling across Carbon Nanotubes in  Far Infrared Terahertz  Electric Fields}

\author{Qi-Lin Zhang$^{1,2}$, Rong-Yao Yang$^1$,  Wei-Zhou Jiang$^{1,}$\footnote{wzjiang@seu.edu.cn},
Zi-Qian Huang$^1$ }
\affiliation{  $^1$ Department of Physics, Southeast University, Nanjing 211189, China\\
$^2$ Department of Mathematics and Physics, Anhui Polytechnic
University, Anhui 241000, China }

\begin{abstract}
  Using molecular dynamics simulations, we investigate
systematically the water permeation properties across the
single-walled carbon nanotube (SWCNT) in the presence of the
terahertz  electric field (TEF). With the TEF normal to the
nanotube, the fracture of the hydrogen bonds results in the giant
peak of net fluxes across the SWCNT with a three-fold enhancement
centered around 14THz. The phenomenon is attributed to the
resonant mechanisms, characterized by librational, rotational, and
rotation-induced responses of in-tube polar water molecules to the
TEF. For the TEF along the symmetry axis of the nanotube, the
vortical modes for resonances and consequently the enhancement of
net fluxes are greatly suppressed by the alignment of polar water
along the symmetry axis, which characterizes the quasi
one-dimensional feature of the SWCNT nicely. The resonances of
water molecules in the TEF can have potential applications in the
high-flux device designs used for various purposes.
\end{abstract}

\maketitle

\section{Introduction}
The study of the water molecular transport in nanochannels is
important  not only for the basic research and potential
applications but also for the understanding of the water
permeation in biological channels. Indeed, one important reason
why research on nanochannel transport properties has continued to
be of interest is that many primary characteristics  of water
confined in simple nanochannels are similar to that of complex
biological
channels~\cite{hu01,wag02,groo01,anis04,beck03,wan05,lot1}.  In
the past, the water transport properties have been studied
extensively in single-walled carbon nanotubes
(SWCNTs)~\cite{mur00,zhu04,gon08,hol06,son08}. Novel studies on
water transports in nanochannels have continued to appear in
recent years with various designs concerning temperature
gradients~\cite{lon07}, charge modification~\cite{li07}, coulomb
dragging~\cite{wang06}, collective mode
oscillations~\cite{zhan13}, and phonon-induced
frictions~\cite{ma15}. Besides the biological implications in
these studies, the stream of these novel progresses may
potentially become integrated into the application pool for the
design of innovative nanofluidic devices, such as flow
sensors~\cite{gho03}, desalination of seawater~\cite{corr08},
molecular sieves~\cite{jira97}, and so on.

Recently, broad interest has been aroused to study the electric
field (EF) effects on the water permeation
~\cite{bon10,rin12,white07,xin06,choi06,jos09,jia11,fig12,li13,kou14,kou15}.
It is known that the applied EF can stem from electromagnetic
radiations. A well-known example is the laser-generated EF that
has fascinating applications, e.g., see~\cite{sala06,fenn10}.
Nowadays, high-power terahertz (THz) radiation can be generated
progressively, say, by optical rectification
methods~\cite{step08}, linear accelerators~\cite{carr02,casa09},
and laser-driven ion or electron
accelerators~\cite{gopa13,pukh13}. The consequent THz EF (TEF) of
the order of GV/m, generated by focusing high-power THz rays, can
open up a new era to study structural transitions in polar
molecules given the TEF-driven molecular rotations over random
thermal motions. Such studies  are of fundamental importance in
revealing the mechanism for molecular motors  that produce
directional motion~\cite{fl05,se11,ti11,zh13}. While for water
permeation in nanochannels, the study of TEF effects is scarcely
found in the literature, and no attention has been paid to the
role of the TEF-driven molecular rotations. It is thus of prime
significance for us to first reveal  whether and how the
TEF-driven molecular rotations and intrinsic mechanisms affect the
water flow.  A clear understanding of the TEF effects is also very
instructive to develop new devices for efficient water filtration
and energy transfer.  Using molecular dynamics (MD) simulations,
we have observed that water permeation properties can greatly be
affected by the TEF through rotational and rotational-induced
translational resonances in a broad THz frequency range.
\section{Computational Methods}
\begin{figure}[thb]
\includegraphics[height=4.5cm,width=8.0cm]{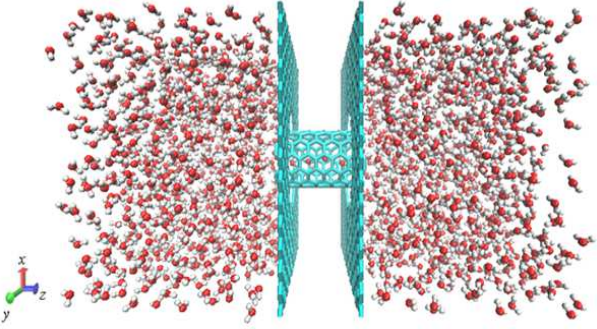}
\caption{(Color online) Snapshot of the simulation system. An
uncapped (6,6) armchair  SWCNT combined with two graphite sheets
is solvated in a water box of
3.5$\times$3.5$\times$6.3\emph{nm}$^{3}$ with 1852 molecules.
There is a vacuum between the two sheets. The axis of SWCNT is
parallel to the z-axis. To obtain the net flux, an acceleration of
each water molecule is imposed, see text. \label{fel}}
\end{figure}

The MD simulation system built with the molecular visualization
program~\cite{hum96} is illustrated in   Fig.~\ref{fel}. An
uncapped (6,6) armchair SWCNT with a length of 1.34 \emph{nm} and
a diameter of 0.81 \emph{nm} is embedded in two parallel graphite
sheets along the $z$ direction. The distance between the bottom
end of the SWCNT and the graphite sheet is 2
\emph{$\overset{\circ}{A}$}. MD simulations are performed using
the NAMD2~\cite{jam05} in a $NVT$ ensemble with the initial box
sizes of $L_x=3.5$ \emph{nm}, $L_y=3.5$ \emph{nm}, $L_z=6.3$
\emph{nm}, constant temperature (300 \emph{K}) achieved by the
Langevin dynamics with a damping coefficient of 1
\emph{ps$^{-1}$}, and periodic boundary conditions in all
directions. The electrostatic interactions are handled by the
particle mesh Ewald method~\cite{dar93} with a real space cutoff
of 1.2 \emph{nm}, and the cutoff for the van der Waals interaction
is also 1.2 \emph{nm}. The CHARMM27 force
field~\cite{mac98,zhu02}, TIP3P water model~\cite{jor83} and SHAKE
algorithm for bond lengths~\cite{ryc77} are adopted in all the
simulations.  The time step is 1 \emph{fs}. To obtain a directed
flow~\cite{zhu02}, an acceleration of 0.01 \emph{nm ps$^{-2}$} per
water molecule is imposed along $+z$ direction to produce a
pressure difference between two ends of the SWCNT about 20
\emph{MPa}~\cite{gon08}. The carbon atoms at the inlet and outlet
are fixed and other atoms of the SWCNT are flexible   to prevent
possible errors from the rigid wall~\cite{gon08,Martini08}.

In our simulations, only the space  between two reservoirs is
exposed to the spatially uniform TEF, and a justification for this
setup is given in the Supporting Material~\cite{zhang}. With the
EF direction being parallel to the $x$-axis, the TEF, acted on the
atomic partial charges, can be expressed as
\begin{equation}\label{eqsym}
\mathbf{E} = E_0({\cos}({\omega}t), 0, 0).
\end{equation}
In this work, we take the amplitude of the EF
$E_0=1\emph{V/nm}$~\cite{luca13,li13} along the $x$-axis, unless
otherwise denoted, and consider  the frequencies ranging from 0.1
to 40 \emph{THz}. Hereafter, these far infrared EFs are still
dubbed the TEF, since the interesting results lie below the region
over a dozen THz. It is known that the electric and magnetic
fields of the electromagnetic wave obey the relation
$|{\bf{E/B}}|=c$ where $c$ is the speed of light. Because the
speed of water molecules inside nanotubes is far less than the
speed of light, the Lorentz force is negligibly small in
comparison to the electric force. Thus, we neglect the effect of
the magnetic field in our estimation.

We should point out that the fast water transport desired in this
work is different from the in-tube water pumping by the AC
electric field~\cite{bon10,rin12}. For water pumping through
SWCNTs, the energy acquisition  in the AC field is necessary, and
simultaneously the spatial asymmetry should be temporally retained
in a delicate thermodynamic relaxation process.
Our simulations with the symmetric setup in the uniform TEF are carefully testified to
have null pumping under zero pressure gradient.

\section{Results and Discussions}

Simulations are performed for the
system within the EF at different frequencies. For each frequency,
the simulation time is 45 ns, and the last 40 ns is collected for
analysis. For clarity, here the net flux is defined  to be the
difference between the water molecular number per nanosecond across
the SWCNT from one end to the other~\cite{gon08,wan05}. The average
occupancy ~\cite{hu01,gon08} and  hydrogen bond numbers of water
molecules inside the SWCNT are denoted by the symbol $\bar{N}$ and
$\bar{N}_{H}$, respectively.

\begin{figure}[thb]
\begin{center}
\vspace*{-5mm}
\includegraphics[height=7cm,width=9cm]{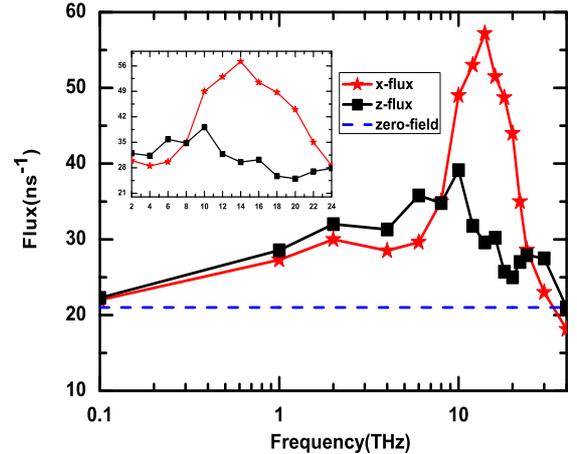}
 \end{center}
\caption{(Color online) The net flux  as a function of the EF
frequency.  The curves with stars and squares stand for results
with the EF along $x$- and $z$-axes, respectively. The dash line
denotes the zero-field flux. The window amplification within 2-24
\emph{THz} is exhibited in the inset. \label{fflux}}
\end{figure}

The net water fluxes, as a function of the EF frequency, are shown in
  Fig.~\ref{fflux} where the zero-field  ($E_0=0$) net flux (about  21
\emph{ns$^{-1}$}) is also displayed. We see that with the EF along
the $x$-axis the net fluxes can be characterized by a giant peak
centered at 14 \emph{THz} with a three-fold enhancement of the
flux.  In sharp comparison to the zero-field result,  the dramatic
enhancement of the water flux is observed in the frequency range
around 10 \emph{THz} to 20 \emph{THz}. For much lower or higher
frequencies, the net flux falls off, being close to the one of the
zero field. For the net fluxes with the EF along the $z$-axis,
also displayed in Fig.~\ref{fflux}, the enhancement is greatly
suppressed. This distinct phenomenon, associated with the
$z$-direction polarization of in-tube water molecules,
characterizes interestingly the property of the quasi
one-dimensional SWCNT.

\begin{figure}[thb]
\vspace*{-10mm}
\includegraphics[height=6cm,width=9cm]{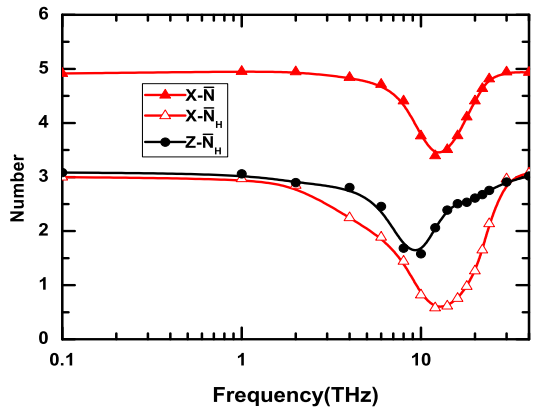}
\caption{(Color online) The average occupancy number
$\bar{N}$ and hydrogen bond number $\bar{N}_{H}$ of water molecules
inside the SWCNT as a function of the EF frequency in the $x$- or
$z$-direction field, as denoted. \label{hhbonds}}
\end{figure}

The extraordinary  characteristics of the water transport should
uniquely be determined by the behaviors of the in-tube water in the
applied TEF. We thus count  the number of hydrogen bonds $\bar{N}_H$
inside the SWCNT in the frequency range as shown in
  Fig.~\ref{hhbonds}, according to the criterion of the oxygen distance
less than 3.5 $\overset{\circ}{A}$ and hydrogen-bond angle $\leq
30^{\circ}$~\cite{wan05}. It is observed that the $\bar{N}_{H}$
values are about  3.01, 2.97, 0.81, 0.61, 1.26 and 2.97 at  0.1,
1, 10, 14, 20 and 30 \emph{THz}, respectively.  The minimum of the
$\bar{N}$ is about 3.5 at 14 \emph{THz}. As shown in
Figs.~\ref{fflux} and \ref{hhbonds}, the deep valley around 14
\emph{THz} in the frequency profile of the $\bar{N}_H$ corresponds
to the giant peak of the flux and the moderate dip of the number
of water molecules in the SWCNT, exhibiting a characteristic
resonant phenomenon. In the whole frequency region interested in
this work, we see that the variation of the flux takes place
coherently with those of the numbers of hydrogen bonds and water
molecules inside the SWCNT. The dramatic drop of the hydrogen-bond
number is a result of the breaking of hydrogen bonds, while the
latter drives the sudden rise of the net flux.

Under the equilibrium condition,  the in-tube water transports in
correlated  bursts of the hydrogen-bonded
filament~\cite{hu01,wag02}, the net flux increases linearly with
the pressure gradient. The water transports at equilibrium can
also be demonstrated by the collective diffusion
model~\cite{zhu04}. Here, the dramatic rise of the net flux occurs
with the rupture of hydrogen bonds due to the energy transfer from
the TEF (see Supporting Material~\cite{zhang}).
The large kinetic energy of water molecules acquired then leads to
fast shuttling of the water molecules from one end to the other
and correspondingly the dramatic rise of the flux. In this case,
the energy transfer from the TEF  spoils the spontaneous diffusion
due to the thermal fluctuation at equilibrium, and we cannot
follow simply the collective diffusion model under the
non-equilibrium condition. In the frequency range without
significant breaking of hydrogen bonds, the fluxes in the presence
of the pressure gradient are consistent with those given by the
approach by Zhu et al~\cite{zhu04}. In the case of dramatic
rupture of hydrogen bonds, the relation between the net flux and
the pressure gradient is not far off linear in a large domain of
the pressure gradient, see Ref.~\cite{zhang}, though in this case
the equilibrium diffusion does not apply  at null pressure.

\begin{figure}[thb]
\begin{center}
\includegraphics[height=5cm,width=7.0cm]{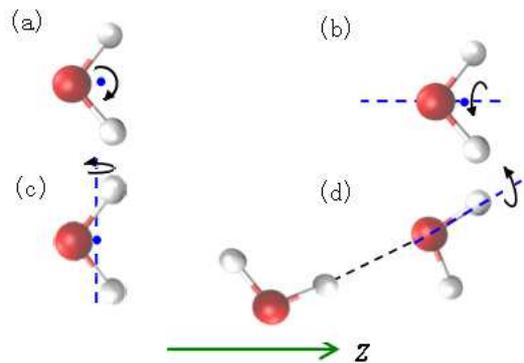}
 \end{center}
\caption{(Color online) The water molecule revolves
along (a), the axis vertical to the water molecule plane through the
center of  mass, (b), the symmetry axis, (c), the axis through the
center of mass but parallel to the connecting line of two hydrogen
atoms,  and (d), the axis on the hydrogen bond,  respectively.
\label{zzhuan}}
\end{figure}

Now, we need to reveal the underlying mechanism for the fracture
of hydrogen bonds. Undoubtedly, the EF can't generate directly the
translational motion of charge neutral molecules, but can drive
the vortical motion of polar water molecules to acquire the
energy. Shown in Fig.~\ref{zzhuan} are four representative
vortical modes with different rotational axes. The vortical motion
can be transmitted to the wobbling of the molecular center of mass
in the vertical section due to the hydrogen-bond locking together
with the thermodynamical motion. As a result, this provides a
pathway to vibrational resonances to shake off the hydrogen
bonding, while the resonant frequency range is estimated around
4-14 \emph{THz}, according to the equality of the energies of the
water molecule binding and the harmonic oscillator~\cite{zhan13}.
Besides the vibrational resonances, the TEF can generate direct
rotational resonances. Due to the hydrogen bonding of water
molecules, all the vortical motions should be, in principle, the
reciprocating motions  rather than the strict rotation. For
convenience, we make somehow discretionary identifications for
vortical modes in Fig.~\ref{zzhuan}:
 ~\ref{zzhuan}(a), (b), and (c) are three independent rotational
modes whose rotational axes are the principal axes, and
 ~\ref{zzhuan}(d) the librational mode for the hydrogen bonding.
Now, we use the rotor model to estimate the resonant frequency for
rotational or librational motions: $E_b=I\omega^2/2$ where $E_b$
is the binding energy of each water molecule, and  $I$ is the
moment of inertia of the water molecule. If $I$ is the total
moment of inertia, the energy of a rotor can be reexpressed by the
components on the three principal axes. Using the value of $E_b=16
$ \emph{kcal/mol}~\cite{hu01}, the resonant frequencies of the
water molecule are estimated to be about 14, 17, 24 and 20
\emph{THz}, corresponding to the modes in   Fig.~\ref{zzhuan}(a),
(b), (c), and (d), respectively. Besides four specific cases, the
axes can tilt for the thermal motion. Consequently, the change of
the moment of inertia results in the continuation to neighboring
frequency domains of the characteristic modes above.

To evidence these resonant mechanisms, we first do a test by
extending the TEF to the reservoirs at both sides. We find nearly
a five-fold enhancement of the net flux in the frequency range of
4-24 \emph{THz}, while beyond this frequency range the flux drops
off rapidly. The reason for this to occur is that the water
molecules liberated from the hydrogen bonding in the reservoir can
break into the nanotube rather freely. In addition, the flux
enhancement is almost independent of the TEF orientation in this
case. As the TEF over the reservoirs withdraws, some distinct
behaviors are exhibited for the single-filed water chain in the
SWCNT. We see from   Fig.~\ref{hhbonds} that the breaking of
hydrogen bonds is not efficient below 8\emph{THz} because of the
SWCNT suppression of the needed large-amplitude vibration. We also
see that the breaking of hydrogen bonds through the rotational
mode of Fig.~\ref{zzhuan}(c) is of low efficiency in that the
coherent flip does take the priority for the constraint of the
SWCNT. In these cases, the suppression of the flux occurs
correspondingly, as shown in Fig.~\ref{fflux}. In the
$x$-direction TEF, an apparent resonant phenomenon, characteristic
of a giant peak, thus arises eventually. We can see that the
resonant frequencies given by the rotor model are consistent with
the large fluxes and small hydrogen bond numbers shown in
Figs.~\ref{fflux} and \ref{hhbonds}. For the $z$-direction TEF, we
see that the rotational mode of Fig.~\ref{zzhuan} (b) is out of
operation. Indeed, all the vortical modes in the $z$-direction TEF
are largely suppressed  by the forced alignment along the
$z$-axis. As a result, the giant peak crashes down, as shown in
Fig.~\ref{fflux}.

In the low-frequency range roughly below 1 \emph{THz}, the
difference from the zero-field flux can be attributed to the
alignment of the polar molecules along the EF. The change of the
alignment direction in the low-frequency TEF yields coherently the
flips of the water chain, resulting in the moderate enhancement of
the flux. With the $z$-direction TEF, the polar water may be
aligned along the $z$-axis, the wobbling and random motions in the
transverse section are largely suppressed with the enhancement of
the ordered permeation. This interprets that the net fluxes remain
a little larger than those of the $x$-direction TEF at low
frequencies. At higher frequencies ($f>30$ \emph{THz}), being far
away from the resonant range, the phenomenon evolves trivially,
being close to the zero-field results.

\begin{figure}[thb]
\begin{center}
\includegraphics[height=6.5cm,width=9.0cm]{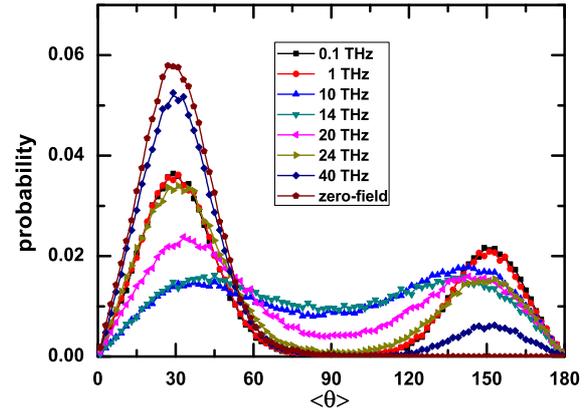}
 \end{center}
\caption{(Color online). Probability distributions of
the dipole  orientation of water molecules inside the tube for
different EF frequencies.\label{hprobability}}
\end{figure}

To comprehend the transport behaviors in the TEF in more details,  we
now examine the average dipole orientations of the water chain.
 Fig.~\ref{hprobability} shows the probability distributions of the
average angle $\langle\theta\rangle$ of water molecules inside the
SWCNT for different frequencies, where $\theta$ represents the angle
between the symmetry axis of the water molecule [see
 Fig.~\ref{zzhuan}(b)] and the $z$-axis. The average is performed over
all the water molecules inside the SWCNT in last 40 \emph{ns} of
the simulation time. As shown in   Fig.~\ref{hprobability}, only
for the zero-field case exists the single peak with the peak value
around $30^{\circ}$, in
accord with previous studies~\cite{jia11, zuo09}.  Note that the symmetric orientation distribution around  $30^{\circ}$ and $150^{\circ}$ at the zero field can be anticipated under small pressure gradients, because in this case low speeds of water molecules can yield sufficient time to produce significant flips in the nanotube.
With the
frequency at 0.1, 1, 24 and 40 \emph{THz} appears remarkably the
double-peak structure, serving as the signature of the frequent
flip of water molecules. The flip may occur for some perturbation
of the form like $e^{i\omega t}$ in a sufficient time span. The
noticeable fact is that the double-peak structure becomes
flattened for the frequencies within the top region of the
resonant peak in
  Fig.~\ref{fflux}, say, 10 , 14 and 20 \emph{THz}, as shown in
  Fig.~\ref{hprobability}. This provides a specific evidence for the
hydrogen bond fracture in the resonant zone. It is the release of
water from the hydrogen bonding that smears out the specific
orientations peaked around $30^{\circ}$ and $150^{\circ}$. Note
that the less flattening for the 20 \emph{THz} curve is also
consistent with the farther distance from the peak value as shown
in   Figs.~\ref{fflux} and \ref{hhbonds}.

In addition, the resonance can be well visualized in a dynamic
evolution of hydrogen bonds for water molecules confined in the
SWCNT.   We observe that the average numbers of hydrogen bonds at
0.1 and 30 \emph{THz} remain almost unchanged from 0 to 10
\emph{ps}. However, we observe that in the frequency range of
10-20 \emph{THz} the hydrogen-bond number starts to reduce
explicitly within 1 ps. These exemplifications indicate that
resonant response of the water chain to the TEF is rather prompt
once the strong torque is exerted on the water molecule.

It is worthwhile discussing the dependence of the water transport
on the TEF amplitude. We find that  the average hydrogen bond
number and the net flux do not undergo significant differences for
$E_0=1$ and a few V/nm.  However, for the $E_0$ below 1
\emph{V/nm}, the TEF effect on the water transport fades away
quickly with the decrease of the TEF amplitude. For instance, at
$E_0=0.1$ \emph{V/nm}, the average hydrogen bond number and the
net flux are similar to their zero-field results, respectively.
The reason of this phenomenon lies in the fact that the
small-amplitude TEF can not produce sufficient torque to win the
competition with the thermal fluctuations of the water molecules.
It is also significant to explore the resonant effects on the
water transport in much longer SWCNTs that is easier to connect
with the experiments. In the SWCNT with the length of  10
\emph{nm}, we have observed the similar rupture of hydrogen bonds.
The flux enhancement due to the rupture of hydrogen bonds is also
about 3-fold at the same pressure gradient as that for the short
SWCNT. Moreover, we  have studied the water transport properties
at  much smaller pressure gradients. It is observed that the
resonant phenomenon is not changed by the magnitude of pressure
gradient. We perform, as an example, the simulation with the
frequencies 10 and 14 \emph{THz}. Compared to net fluxes at
non-resonant frequencies, the significant flux enhancement remains
at lower pressures. For details, see the Supporting
Material~\cite{zhang}.

We notice that the TEF should be regarded as the effective field
of the THz radiation penetrating the tube because of the partial
screening by the electron clouds laying on the SWCNT. It is of
interest to consider later on whether the radial breathing mode
(RBM) of the SWCNT can be efficiently generated by the TEF and how
RBM-induced vibrations mix with the direct TEF-driven
modes~\cite{zhan13}. In addition, we would mention that
simulations with polarizable models for conducting carbon
nanotubes do not produce appreciable difference in the water
transport, compared with those with nonpolarizable models for
semiconducting carbon nanotubes~\cite{moul05, jose08}. Finally, in
order to examine the model dependence, we performed some
comparative simulations using the SPC/E water
model~\cite{ber87,mark01}. Though the magnitude of the flux with
the TIP3P water model is different from the one obtained with the
SPC/E model,  the tendencies of the net flux, the hydrogen-bond
and occupancy numbers with both models are consistent with each
other. Thus, the main conclusions that we have drawn do not depend
on the specific water model. For details, see Ref.~\cite{zhang}.

\section{Conclusions}
In summary, we have demonstrated the effects of the vortical
resonant mechanisms in the TEF on the transport properties and
dynamical behaviors of water molecules across the SWCNT. The
resonances, resulting from librational, rotational and
rotation-induced translational modes of polar water molecules,
generate significant hydrogen-bond fractures and consequently the
dramatic enhancement of the net flux, with the TEF normal to the
SWCNT. The resonant phenomenon is well established.   The change
shows up when the TEF turns to the symmetry axis of the nanotube:
the resonant phenomenon is greatly suppressed by the alignment of
water molecules along the symmetry axis, which characterizes the
quasi one-dimensional feature of the SWCNT. Our work suggests that
the TEF resonant mechanisms are instructive to develop high-flux
nanoscale devices and explore possible implications for the
biological water channelling.

\section*{Acknowledgement}

We thank Profs. Chang-Bo Fu,  Zhen-Hua Ni, Ai-Guo Li, Drs. Hua Chen, Kun-Quan
Hong, Jian Liu, Dong-Rui Zhang and  Sina Wei for useful discussions and Dr.
Ren-De Miao for computational help. The work was supported in part by
the National Natural Science Foundation of China under Grants Nos.
10975033, and 11275048, the China Jiangsu Provincial Natural Science
Foundation under Grant No.BK20131286, and Natural Science Foundation
of Anhui Province of China under Grant No.1508085QA19.

\end{document}